\documentclass[aps,prb,twocolumn,a4paper,superscriptaddress]{revtex4-1}
\usepackage{graphicx,amsmath,amssymb,mathrsfs,array,longtable,delarray,color}

\begin{document}
\newcommand{\VGoneDC}{$V_{\text{G1}}^{\text{dc}}$}
\newcommand{\VGoneRF}{$V_{\text{G1}}^{\text{rf}}$}
\newcommand{\VGtwo}{$V_{\text{G2}}$}
\newcommand{\VGthreeDC}{$V_{\text{G3}}^{\text{dc}}$}
\newcommand{\VGthreeRF}{$V_{\text{G3}}^{\text{rf}}$}
\newcommand{\dIbydV}{$dI_{\text{T}} / dV_{\text{G3}}^{\text{dc}}$}
\newcommand{\dIbydtau}{$dI_{\text{T}} / d\tau_{\text{d}}$}
\newcommand{\It}{$I_{\text{T}}$}
\newcommand{\Ip}{$I_{\text{P}}$}
\newcommand{\Ir}{$I_{\text{R}}$}
\newcommand{\tauD}{$\tau_{\text{d}}$}

%\preprint{}

\title{Measurement and control of electron wave packets from a single-electron source}
\date{July 22, 2015}


\author{J.~Waldie}
%\thanks{}
\affiliation{National Physical Laboratory, Hampton Road, Teddington, Middlesex TW11 0LW, UK}
\affiliation{Cavendish Laboratory, University of Cambridge, J.J.~Thomson Avenue, Cambridge CB3 0HE, UK}
\author{P.~See}
\affiliation{National Physical Laboratory, Hampton Road, Teddington, Middlesex TW11 0LW, UK}
\author{V.~Kashcheyevs}
\affiliation{Faculty of Physics and Mathematics, University of Latvia, Zellu Street 8, LV-1002, Riga, Latvia}
\author{J.P.~Griffiths}
\author{I.~Farrer}
\author{G.A.C.~Jones}
\author{D.A.~Ritchie}
\affiliation{Cavendish Laboratory, University of Cambridge, J.J.~Thomson Avenue, Cambridge CB3 0HE, UK}
\author{T.J.B.M.~Janssen}
\author{M.~Kataoka}
\affiliation{National Physical Laboratory, Hampton Road, Teddington, Middlesex TW11 0LW, UK}


\date{\today}

\begin{abstract}
We report an experimental technique to measure and manipulate the arrival-time and energy distributions of electrons emitted from a semiconductor electron pump, operated as both a single-electron source and a two-electron source. Using an energy-selective detector whose transmission we  control on picosecond timescales, we can measure directly the electron arrival-time distribution and we determine the upper-bound to the distribution width to be 30~ps. We study the effects of modifying the shape of the voltage waveform that drives the electron pump, and show that our results can be explained by a tunneling model of the emission mechanism. This information was in turn used to control the emission-time difference and energy gap between a pair of electrons.
\end{abstract}

%\keywords{}

\maketitle


%\subsection{}
%\subsubsection{}

\section{Introduction}
The ability to emit, coherently control and detect single electrons is highly desirable for quantum information processing applications \cite{Loss1998, Barnes2000} and experiments exploring the fermionic quantum behavior of electrons. Semiconductor two-dimensional electron systems (2DES) in perpendicular magnetic fields offer the possibility of ballistic, coherent electron transport over tens of microns \cite{Roulleau2008} in chiral one-dimensional (1D) quantum Hall edge channels,\cite{Halperin1982} the electronic equivalent of the fiber-optic for photons. Several experiments have used electrons in 1D edge channels, with quantum point contacts as electron beam splitters, to perform electron-quantum-optics-type experiments. \cite{Henny1999, Oliver1999, Ji2003} The realization of triggered single-electron emitters in semiconductors \cite{Kouvenhoven1991, Fujiwara2004, Blumenthal2007, Kaestner2008, Feve2007} allows these experiments to be performed using single-particle states. Using a mesoscopic capacitor \cite{Feve2007} as a source of single electron-hole pairs, Bocquillon \textit{et al} performed noise-correlation measurements using the Hanbury Brown and Twiss \cite{Bocquillon2012} and Hong-Ou-Mandel \cite{Bocquillon2013} geometries. In these experiments, the single particles emitted by the mesoscopic capacitor lie close to the Fermi energy. \cite{Bocquillon2012, Parmentier2012} In contrast, the tunable-barrier quantum dot electron pump \cite{Blumenthal2007, Kaestner2008} can inject single electrons into edge states more than 100 meV above the Fermi level. \cite{Fletcher2013, Leicht2011} The high electron energy limits the mixing of the emitted electrons with the low-energy Fermi sea, and this enabled Fletcher \textit{\textit{et al}} to measure the emitted electron wavepackets, distinct from the Fermi sea, with temporal resolution of $\sim80$~ps. \cite{Fletcher2013} Using a similar device geometry, Ubbelohde \textit{et al} measured the partitioning noise of electron pairs from an electron pump incident on an electronic beam splitter, revealing regimes of independent, distinguishable or correlated partitioning, with the origin of the latter not yet understood. \cite{Ubbelohde2015} These results call for more detailed studies of the electron emission process with higher temporal resolution, with a view to controlling the emitted wavepackets.

Here we study the arrival-time and energy distributions of electrons emitted by a single-electron pump, using an energy-selective detector \cite{Fletcher2013, Palevski1989, Sivan1989} which we control on picosecond timescales using an arbitrary waveform generator (AWG). This time-resolution allows us to measure directly the time distribution of electrons arriving at the detector, which we find to have a full-width-at-half-maximum (FWHM) of 30~ps or less. We also use the AWG to engineer the ac voltage waveform driving the pump, \cite{Giblin2012} so as to study the link between the electron emission process and the shape of the driving waveform. We observe distinct features in the electron energy distribution linked to the digital nature of the waveform, which we explain using a tunneling model of the emission process. \cite{Kashcheyevs2010, Kashcheyevs2012} Using these insights, we demonstrate manipulation of the electron emission by operating the pump as a two-electron source and modifying the emission-time difference and energy gap between the two electrons.

\section{Experimental Methods}
\begin{figure}
\centering
\includegraphics[width=0.5\textwidth]{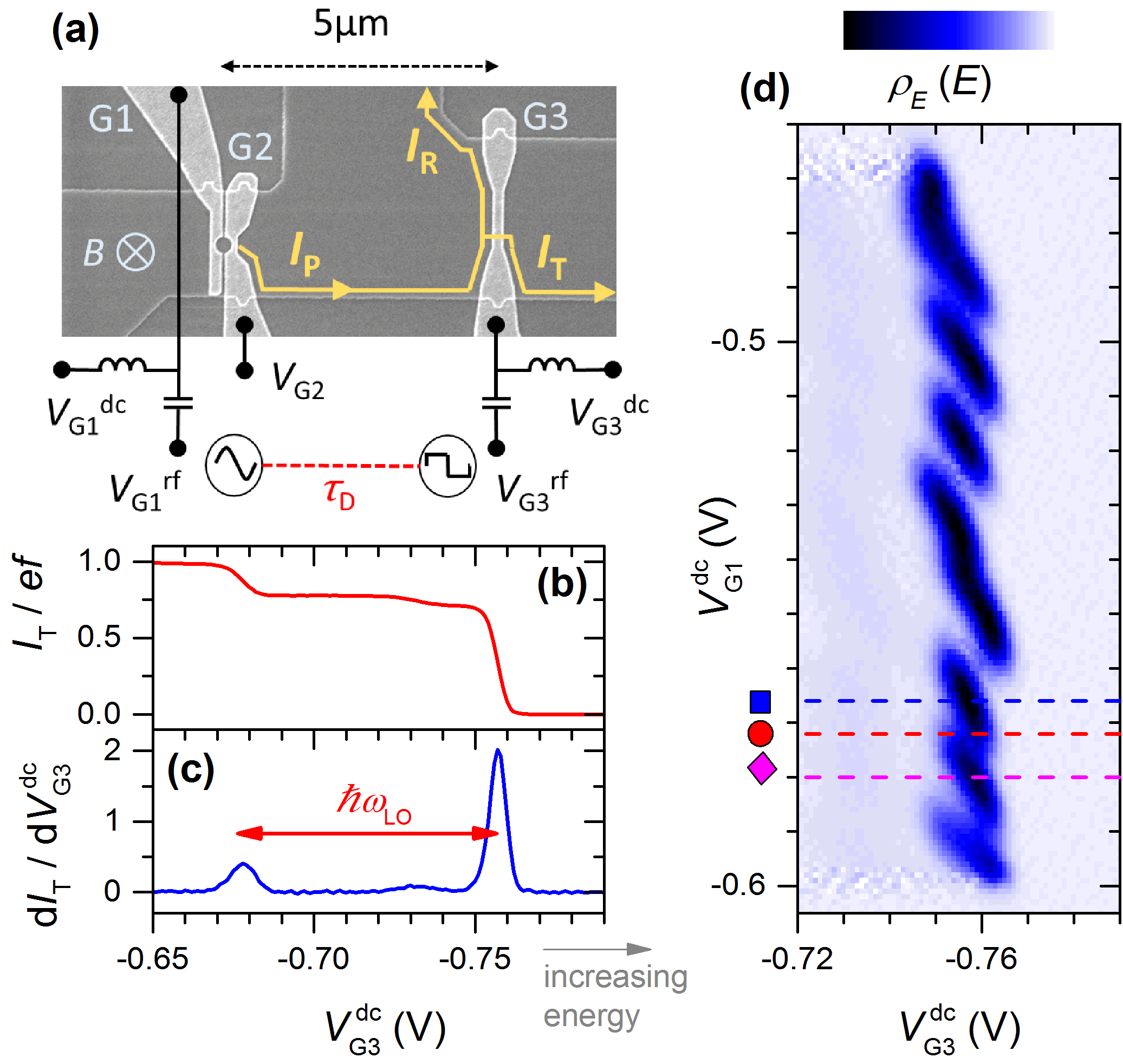}
\caption{(a) Scanning electron micrograph of the device, indicating the measured currents and applied voltages. (b) \It\ as a function of the static detector voltage \VGthreeDC. (c) Derivative \dIbydV, which is proportional to the electron energy distribution $\rho_E(E)$. (d) Evolution of $\rho_E(E)$ with \VGoneDC. Data taken at \VGtwo\ $=-0.535$~V and $B=14$~T. In (b) and (c), $\text{\VGoneDC}=-0.566$~V, corresponding to the upper dashed line in (d). Colored dashed lines and symbols (square, circle, diamond) in (d) indicate values of \VGoneDC\ that we study in Fig.~\ref{fig:TimeDist}(a)-(c). (Color online)}
\label{fig:EnergyDist}
\end{figure}

Our device consists of a tunable-barrier quantum dot electron pump \cite{Blumenthal2007, Kaestner2008, Giblin2012} and a tunable potential barrier detector, \cite{Fletcher2013} defined in a 2DES in a GaAs/AlGaAs heterostructure [see Fig.~\ref{fig:EnergyDist}(a)]. Negative bias voltages \VGoneDC\ and \VGtwo\ applied to gates G1 and G2 define a quantum dot region between entrance and exit barriers. The entrance gate voltage is modulated with an ac waveform \VGoneRF\ of frequency $f = 120$~MHz and peak-to-peak amplitude 1~V, from a two-channel AWG. \footnote{Tektronix AWG7122C} This modulation periodically lowers the entrance barrier, so that electrons tunnel from the left reservoir into the dot, and then raises the dot potential so that the electrons tunnel out to the right reservoir. An integer number, $n$, of electrons is pumped per cycle, resulting in dc pumped current \Ip$=nef$, where $e$ is the electron charge. We note that the waveform produced by the AWG is a digital reconstruction of a sinusoidal wave, with sampling rate 12~GS/s and analog bandwidth 5~GHz, and is transmitted to the sample via 50~$\Omega$ impedance co-axial cables and a bias-tee of bandwidth 12~GHz. Measurements are carried out at 300~mK in a perpendicular magnetic field of $10-14$~T (corresponding to Landau level filling factors $\nu<1$ in the bulk 2DES). Due to the magnetic field applied in the direction shown in Fig.~\ref{fig:EnergyDist}(a), pumped electrons travel in edge states to the detector barrier (G3), with pump-to-detector distance 5~$\mu$m. At the detector they are either transmitted or reflected, depending on the electron energy $E$ relative to the detector barrier height $E_D = E_o -\beta V_\text{G3}$ ($E_o$ and $\beta$ are constants), giving dc transmitted and reflected currents \It\ and \Ir. For a sufficiently low detector barrier height, we find $\text{\It}\approx\text{\Ip}$ and $\text{\Ir}\approx0$, \cite{Fletcher2013} as expected for chiral edge state transport. From measurements of \It\ we determine the time and energy distributions for electrons arriving at the detector.

With constant detector voltage $V_\text{G3} = \text{\VGthreeDC}$, assuming the detector transmission $T(E-E_D)$ is 1(0) for $E>E_D$ ($E<E_D$), we can estimate the energy distribution from $\rho_E(E) \propto \text{\dIbydV}$ (corrections to this approximation will be discussed later). To investigate the time distribution, we add a 120~MHz square wave $\text{\VGthreeRF}(t)$, with peak-to-peak amplitude 32~mV, to the detector, with a controllable time delay \tauD\ relative to the pump drive waveform, giving $V_\text{G3} = \text{\VGthreeDC} + \text{\VGthreeRF}(t+\text{\tauD})$. The square wave is generated by the second channel of the AWG and is transmitted to the sample via 50~$\Omega$ impedance co-axial cables and a bias-tee of bandwidth 6~GHz. For suitable values of \tauD\ the electron wave packet will arrive at the detector just as the square wave \VGthreeRF\ changes from positive to negative and (for suitable \VGthreeDC) the detector transmission probability switches rapidly from 1 to 0, so only the fraction of the electron wave packet arriving before the switch will be transmitted. For a perfectly sharp switch, the arrival-time distribution is given by $\rho_t(t) \propto \text{\dIbydtau}$. A similar technique has recently been used to measure the time-of-flight of edge magnetoplasmons in a 2DES. \cite{Kamata2010, Kumada2011} This method of estimating $\rho_t(t)$ has advantages compared to the method previously applied to pumped electrons in Ref.~\onlinecite{Fletcher2013}, where the detector modulation was sinusoidal and the arrival-time distribution was deduced from changes in the apparent energy broadening with \tauD. First, our method gives a more direct measurement of $\rho_t(t)$ and, second, it allows us to use a much smaller detector modulation amplitude, reducing back-action of the detector on the electron pump.

\section{Energy Distribution}
In Fig.~\ref{fig:EnergyDist}(b)-(d) we present measurements of the electron energy distribution for single-electron pumping, where \Ip\ $= ef \approx 19$ pA. Figure \ref{fig:EnergyDist}(b) is a typical plot of the transmitted current \It\ as a function of \VGthreeDC\ for a static detector (\VGthreeRF\ = 0). As the detector barrier is raised, \It\ decreases from $\approx\text{\Ip}$ to 0, in two main steps (there is an additional small step barely visible around \VGthreeDC $=-0.73$~V, but we do not yet know the origin of this step \footnote{This small step corresponds to an additional small peak in the energy distribution, about 10~meV below the main peak. This additional peak was always found $\approx10$~meV below both main peaks, independent of \VGoneDC\ and \VGtwo, and of the applied magnetic field.}). The corresponding energy distribution, estimated from \dIbydV, has two main peaks [Fig.~\ref{fig:EnergyDist}(c)]. We use the method of Taubert \textit{et al} \cite{Taubert2011} to determine the conversion factor between \VGthreeDC\ and electron energy, $dE/d$\VGthreeDC\ $\approx-(0.50\pm0.05)e$. The separation between the peaks in the energy distribution is $\approx40$ meV, consistent with the longitudinal optic (LO) phonon energy 36 meV in GaAs \cite{Blakemore1982, Taubert2011} within the uncertainty of our energy conversion factor. Therefore we attribute the lower energy peak to electrons that have emitted an LO phonon. We find that the probability of phonon emission decreases as $B$ is increased from 8 T to 14 T, as observed by Fletcher \textit{et al}. \cite{Fletcher2013} In the following, we focus on the higher-energy peak, due to the electrons that do not emit phonons, which has FWHM $\approx3.5$ meV. This energy spread reflects not only the electron emission energy distribution, but also several types of experimental energy broadening, which will be discussed later.

The electron energy distribution can be varied by changing the pump entrance and exit gate voltages. \cite{Leicht2011} As observed by Fletcher \textit{et al}, \cite{Fletcher2013} we find that the energy distribution shifts linearly to higher energy as the exit gate voltage \VGtwo\ is made more negative. This is because with a higher exit barrier the electrons require more energy to tunnel out of the pump. However, we see a very different dependence on the entrance gate voltage \VGoneDC, as shown in Fig~\ref{fig:EnergyDist}(d). The total entrance gate voltage is the sum $V_\text{G1} = \text{\VGoneDC}+\text{\VGoneRF}$. We might expect emission to occur when the total voltage $V_\text{G1}$ reaches a certain threshold value. In this case a shift $\Delta\text{\VGoneDC}$ would shift the emission time by $-\Delta\text{\VGoneDC}/(d\text{\VGoneRF}/dt)$, but not the emission energy. In contrast to this simple picture, Fig.~\ref{fig:EnergyDist}(d) shows that the peak in the energy distribution follows a series of diagonal lines as a function of \VGoneDC. As \VGoneDC\ becomes more negative, the peak in $\rho_E(E)$ shifts to higher energy and then diminishes in amplitude, being replaced by a new peak at lower energy. In the following sections, we show how these features arise from the details of the pumping waveform, giving insight into the electron emission process.

\section{Time Distribution}
\begin{figure}
\centering
\includegraphics[width=0.5\textwidth]{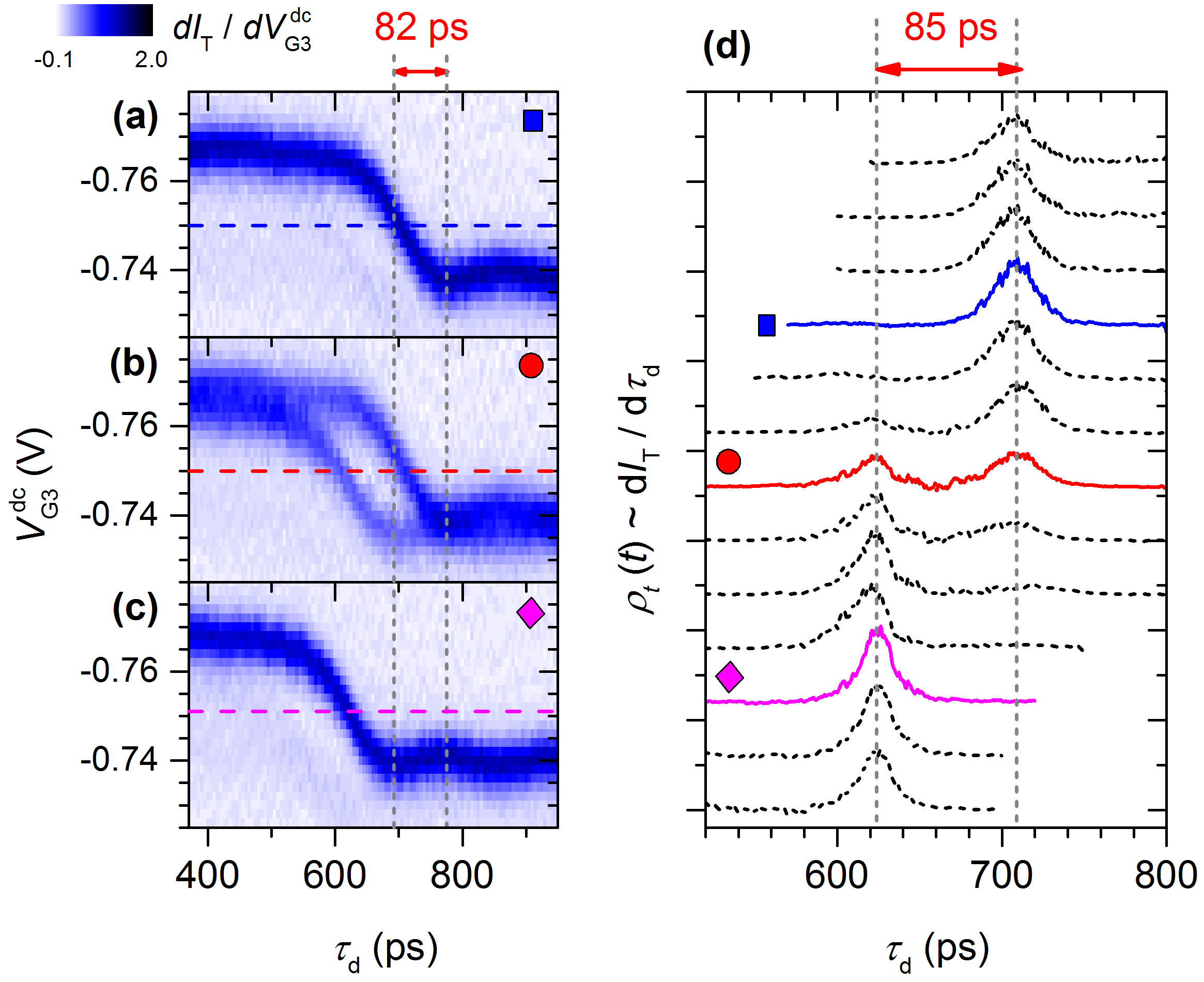}
\caption{(a)-(c) \dIbydV\ as a function of \tauD\ at the three values of \VGoneDC\ indicated by dashed lines and symbols in Fig.~\ref{fig:EnergyDist}(d); (a) 
$\text{\VGoneDC} = -0.566$~V, (b) $\text{\VGoneDC} = -0.572$~V, (c) $\text{\VGoneDC} = -0.580$~V; the approximate shift of $\Delta\text{\tauD}=82$~ps between the pattern in (a) and (c) is indicated; this shift is approximately equal to the AWG sampling interval of 83.3~ps. (d) $\rho_t(t)$, estimated from \dIbydtau, as we change \VGoneDC\ from (top) $-0.560$~V to (bottom) $-0.584$~V in steps of 2 mV. Each trace in (d) is measured at the \VGthreeDC\ that ensures the electron energy distribution is centered between the high and low values of the modulated detector barrier height [dashed lines in (a)-(c) indicate the \VGthreeDC\ used]; colored solid traces and symbols in (d) correspond to the same \VGoneDC\ as in (a) to (c). Data taken at \VGtwo\ $=-0.535$~V and $B=14$~T. (color online)}
\label{fig:TimeDist}
\end{figure}
To gain understanding of the behavior in Fig.~\ref{fig:EnergyDist}(d), we study the electron arrival-time distribution using the square-wave detector modulation. In Fig.~\ref{fig:TimeDist}(a)-(c) we show how the derivative \dIbydV\ changes as we sweep the time delay \tauD\ of the detector square wave \VGthreeRF, for three different values of \VGoneDC\ [corresponding to colored symbols in Fig.~\ref{fig:EnergyDist}(d)]. This derivative is no longer a simple measure of the energy distribution, because the transmitted current now depends on whether the electrons arrive at the detector when \VGthreeRF\ is high or low. For small(large) \tauD\ the peak in \dIbydV\ is shifted to more negative(positive) \VGthreeDC\ because the electrons arrive at the detector in the positive(negative) half of the square wave. The position (in \tauD) of the crossover between the two regimes indicates the electron arrival time at the detector (plus a constant offset due to different propagation lengths for the two ac signals). From the horizontal shift between the patterns of Fig.~\ref{fig:TimeDist}(a) and (c), we see that the electron arrival is shifted earlier in time by approximately 82~ps as we change \VGoneDC\ from $-0.566$~V to $-0.580$~V. However, the peak in the time distribution does not shift continuously with \VGoneDC. At intermediate \VGoneDC\ ($-0.572$~V) the arrival-time distribution is split into two [Fig.~\ref{fig:TimeDist}(b)]. We note from Fig.~\ref{fig:EnergyDist}(d) that the energy distribution is also bimodal at this \VGoneDC.

In Fig.~\ref{fig:TimeDist}(d) we present the arrival-time distribution $\rho_t(t)$, estimated from \dIbydtau, as we vary \VGoneDC\ from (top) $-0.560$~V to (bottom) $-0.584$~V in steps of 2 mV. Each trace is taken by sweeping \tauD\ at constant \VGthreeDC, being careful to choose \VGthreeDC\ such that the entire electron energy distribution is between the high and low values of the detector barrier. For \VGoneDC\ $=-0.560$~V (top trace) the arrival-time distribution has a single peak, centered at \tauD\ $\approx709$~ps. As \VGoneDC\ becomes more negative the time distribution remains constant, although we know from Fig.~\ref{fig:EnergyDist}(d) that the energy distribution shifts to higher energy. However, at \VGoneDC\ $\sim-0.570$~V this peak in $\rho_t(t)$ weakens and a new peak emerges at \tauD\ $\approx624$~ps (85~ps earlier), which dominates the distribution for \VGoneDC\ $<-0.574$~V. In the same voltage range, the peak in $\rho_E(E)$ is replaced by a lower energy peak. This behavior is periodic in \VGoneDC, with the peaks in $\rho_t(t)$ and $\rho_E(E)$ being replaced by new peaks at earlier time and lower energy, roughly every 15 mV. 

The narrowest time distribution in Fig.~\ref{fig:TimeDist}(d) has FWHM $\approx30$~ps, significantly narrower than the 80~ps result of Fletcher \textit{et al}. \cite{Fletcher2013} Thus we have achieved improved time resolution in the measurement of the electron wave packet emitted by an electron pump. The improvement in resolution comes from modulating the detector barrier with a square wave, giving faster switching of the barrier height from high to low. From Fig.~\ref{fig:TimeDist}(a), we estimate the maximum rate of change $dV_\text{G3}/dt\approx 0.26$~mV/ps, nearly four times faster than in Ref.~\onlinecite{Fletcher2013}. However, this rate is still finite and, combined with the width of the electron energy distribution (3.5 meV), gives the main limitation to our time resolution. Electrons with different energies are reflected/transmitted for slightly different \tauD, broadening the measured time distribution. Therefore we believe that 30~ps is likely to be an over-estimate of the true wave packet width.

\section{Emission Mechanism}
The spacing of 85~ps between the peaks in the time distribution of Fig.~\ref{fig:TimeDist}(d) is close to the AWG sampling interval, $(12\text{~GHz})^{-1}=83.3$~ps, suggesting that electron emission may be influenced by the digital nature of the pumping waveform \VGoneRF. The AWG generates a waveform of frequency 120~MHz by cycling through a list of 100 voltage values at 12~GS/s, with the voltage updated once every 83.3~ps. The details of the waveform reaching the device depend on the limited-bandwidth frequency response of the signal line, which includes co-axial conductors, connectors, attenuators and a bias-tee. Measurements of the AWG waveform with an oscilloscope of sampling rate 60 GS/s, using the same room temperature co-axial cables and bias-tee, show a pronounced quasi-sinusoidal ripple of frequency 12~GHz and a peak-to-peak amplitude 14~mV superimposed on the intended 120-MHz sine wave. The ripple frequency matches the AWG sampling rate 12~GS/s. In this section, we use a simple model of the electron emission process to show that each of the peaks in the electron arrival-time and energy distributions is due to electrons being emitted in different sampling intervals of the AWG waveform. We note that the square wave signal applied to our detector does not seem to show a significant 12-GHz ripple, probably because a bias-tee of bandwidth 6~GHz was used for this signal.

\begin{figure}
\centering
\includegraphics[width=0.5\textwidth]{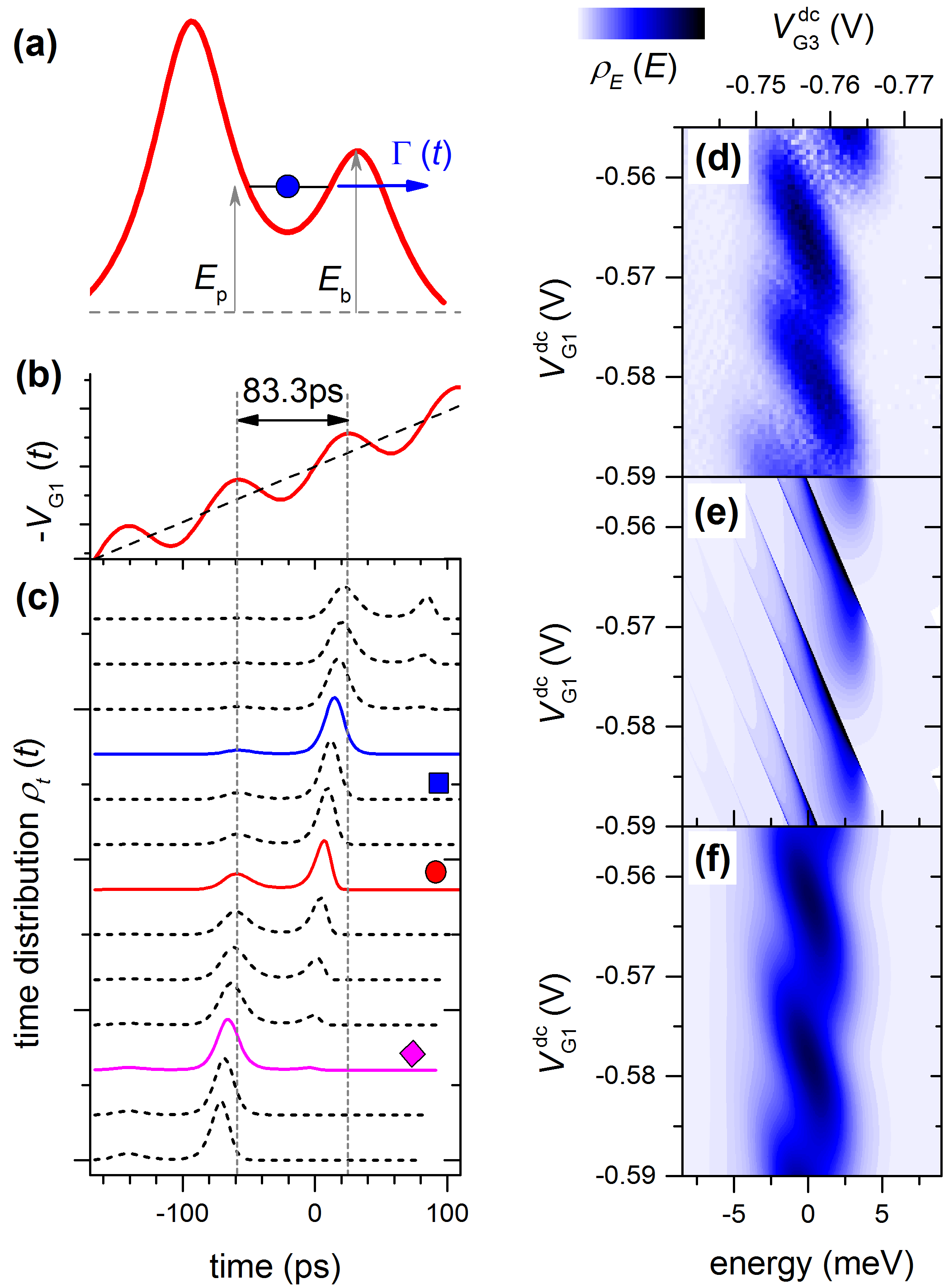}
\caption{(a) Electron potential profile in the electron pump at the point of electron emission. (b) Model of $V_\text{G1}(t)$ close to the emission point; the ideal 120~MHz waveform rises approximately linearly (dashed line) but the AWG adds a 12-GHz ripple (solid line). (c) Modeled $\rho_t(t)$ at the same values of \VGoneDC\ shown in Fig.~\ref{fig:TimeDist}(d); colored solid traces and symbols indicate the correspondence. (d)-(f) $\rho_E(E)$ as a function of \VGoneDC\ (d) measured $\rho_E(E)$, (e) modeled $\rho_E(E)$ without accounting for experimental broadening, (f) model including broadening due to the energy-dependence of the detector barrier transmission. (color online)}
\label{fig:Modelling}
\end{figure}
Figure \ref{fig:Modelling}(a) illustrates the potential profile for the electron bound in the dynamic quantum dot of the pump just before the electron is emitted. Following the approach of Refs.~\onlinecite{Kashcheyevs2010, Kashcheyevs2012}, which describe the ``back-tunneling'' of electrons through the entrance barrier just after electrons are loaded into the pump, we model the electron emission process of ``forward-tunneling'' through the exit barrier. We approximate the time-dependent entrance gate voltage close to the emission point as [see Fig.~\ref{fig:Modelling}(b)]
\begin{equation}
V_\text{G1}(t) = \text{\VGoneDC} - \lvert\dot{V}_\text{G1}\rvert t+ V_\text{G1}^\text{12G}\sin(2\pi f_s t),
\label{eqn:VG1}
\end{equation}
where $-\lvert\dot{V}_\text{G1}\rvert$ is the rate of change of the ideal 120~MHz sinusoidal waveform close to the emission point and $V_\text{G1}^\text{12G}$ is the amplitude of the 12-GHz ripple. The rate of electron tunneling through the exit barrier is
\begin{equation}
\Gamma(t) = \Gamma_0 \exp\left[-\frac{E_b(t)-E_p(t)}{\Delta_b} \right],
\label{eqn:Gamma}
\end{equation}
where $E_b(t)$ and $E_p(t)$ are the exit barrier height and the electron energy level in the pump, and $\Delta_b$ depends on the shape of the exit barrier. \cite{Kashcheyevs2012} Equation (\ref{eqn:Gamma}) is valid provided that $\Gamma(t)\ll\Gamma_0$, i.e. electron emission is by tunneling, rather than ballistic. \cite{Ubbelohde2015} We assume that the lever-arm factors $\alpha_{b(p)} = -dE_{b(p)}/dV_\text{G1}$ are frequency-independent. Then we can re-write Eqn.~(\ref{eqn:Gamma}) as
\begin{equation}
\Gamma(t) = \frac{1}{\tau}\exp\left[ \frac{t-t_e}{\tau} + \frac{A}{2\pi f_s \tau}\sin(2\pi f_s t) \right].
\label{eqn:lnGamma}
\end{equation}
Here, $\tau^{-1} = (\alpha_p-\alpha_b)\lvert\dot{V}_\text{G1}\rvert/\Delta_b$,
$A=2\pi f_s V_\text{G1}^\text{12G}/\lvert\dot{V}_\text{G1}\rvert$ and
$t_e = \text{\VGoneDC}/\lvert\dot{V}_\text{G1}\rvert+\text{const}$. Similarly, the electron energy level $E_p(t)$ can be written as
\begin{equation}
E_p(t) = E_0 + \Delta_\text{ptb}\left[ \frac{t-t_e}{\tau} + \frac{A}{2\pi f_s \tau}\sin(2\pi f_s t) \right],
\label{eqn:Energy}
\end{equation}
where $\Delta_\text{ptb}$ is the ``plunger-to-barrier ratio'', $\alpha_p\Delta_b/(\alpha_p-\alpha_b)$. \cite{Kashcheyevs2012,Kaestner2014}
From Eqn.~(\ref{eqn:lnGamma}) and the rate equation $dp(t)/dt = -\Gamma(t)p(t)$ we calculate the probability $p(t)$ that the electron remains in the pump at time $t$, and the corresponding emission-time distribution $\rho_t(t) = -dp/dt$. Here, for the sake of simplicity, we assume the measured arrival-time distribution to be the equal to the emission-time distribution; we neglect the electron dispersion and the time of flight between the pump and the detector, which will be the subject of a future publication. Also, we assume that the energy of an electron arriving at the detector is the same as the energy level $E_p(t)$ at the time of emission, so we find the energy distribution $\rho_E(E)$ from $\rho_t(t)$ and Eqn.~(\ref{eqn:Energy}). \cite{Leicht2011}

We estimate the amplitude of the 12-GHz ripple to be $V_\text{G1}^\text{12G}\approx 7$ mV, based on the oscilloscope measurements of the AWG waveform mentioned previously. The typical slope of the programmed waveform in the emission region is $\lvert\dot{V}_\text{G1}\rvert = 16$~mV per sampling interval, and we have separately estimated the lever-arm factor $\alpha_p \approx 0.25e$. These values give $A\approx2.75$ and $\Delta_\text{ptb}\approx1$ meV. Therefore $\tau$ is the only adjustable parameter in our model, apart from additive constants. We note that $\tau$ is the characteristic timescale over which the integrated tunneling rate [Eqn.~(\ref{eqn:Gamma})] becomes large, so we expect $\tau$ to be comparable to the wave packet width in the time domain.

Figure \ref{fig:Modelling}(c) presents the modeled emission-time distribution $\rho_t(t)$ for the same values of \VGoneDC\ as in Fig.~\ref{fig:TimeDist}(d), using $\tau=(4f_s)^{-1}\approx20$~ps. The modeled $\rho_t(t)$ shows a series of peaks with separation $\sim(f_s)^{-1} \sim83$~ps, with gradual shift in weight to earlier-time peaks as \VGoneDC\ becomes more negative. The peaks in $\rho_t(t)$ come at, or just before, the local maxima in $- V_\text{G1}(t)$, where the emission rate $\Gamma(t)$ is also maximized. Thus the approximately constant peak positions in the time distribution of Fig.~\ref{fig:TimeDist}(d) arise because the tunneling rate does not rise monotonically with time but has a series of local maxima, approximately 83~ps apart. The experimentally measured separation between the peaks in $\rho_t(t)$ (85~ps) differs slightly from this, probably because the ripple in the AWG waveform is only quasi-periodic.

The measured and the modeled energy distributions $\rho_E(E)$ for a range of \VGoneDC\ are shown in Fig.~\ref{fig:Modelling}(d) and (e). As in the experimental results, each peak in the modeled $\rho_E(E)$ shifts linearly towards higher energy as \VGoneDC\ is made more negative, then gradually fades and is replaced by another peak at lower energy. The linear shift occurs because emission is concentrated around the local maxima in $-V_\text{G1}(t)$ and the energy $E_p$ at these times increases as we make \VGoneDC\ more negative. Emission only shifts to an earlier, lower-energy local maximum for a sufficient change in \VGoneDC. However, although the model reproduces the positions of the peaks in $\rho_E(E)$, it predicts a rather different peak shape from the approximately Gaussian peaks in the measured energy distribution. The modeled peak shape is narrower, and shows a sharp peak on the higher energy side. This sharp peak corresponds to electron emission at the local maximum in $-V_\text{G1}(t)$, where $dE_p(t)/dt = 0$ (a smaller sharp peak occurs due to the small amount of emission at the local minimum). We suggest these sharp peaks in $\rho_E(E)$ are not observed experimentally due to several factors that broaden the measured energy distribution, including the energy-dependence of the detector barrier transmission, gate voltage noise and inelastic scattering. These broadening mechanisms are not easy to distinguish from one another experimentally. We include such broadening by modeling the barrier transmission $T$ as a non-ideal step function
\begin{equation}
T(E) = \frac{1}{1+\exp\left[-(E-E_D)/\Delta_D\right]},
\label{eqn:Transmission}
\end{equation}
where $E_D = -0.5e\text{\VGthreeDC}+\text{const.}$ is the height of the detector barrier and $\Delta_D$ quantifies the broadening. Using Eqn.~(\ref{eqn:Transmission}) and the model energy distribution of Fig.~\ref{fig:Modelling}(e), we find the energy distribution that would be measured from \dIbydV, with results shown in Fig.~\ref{fig:Modelling}(f). The inclusion of broadening gives much better agreement with the experimental results and for $\Delta_D=0.8$ meV the modeled peak width matches the experimental value. Therefore we believe this simple model can explain the key features of our observations.

For the model results in Fig.~\ref{fig:Modelling} we have assumed the characteristic tunneling time-scale $\tau\approx20$~ps. For much larger $\tau$, the peaks in the modeled $\rho_t(t)$ become too broad to be consistent with the measured peak width ($\sim30$~ps). On the other hand, for $\tau\ll20$~ps the electron emission shifts from the local maxima in $- V_\text{G1}(t)$ to the risers before the maxima, and this destroys the linear dependence of the peaks in $\rho_E(E)$ on \VGoneDC. Therefore a value of $\tau\approx20$~ps is most consistent with our experimental observations.

We note that our observation of a 30~ps wave packet width may be specific to the details of the AWG pumping waveform that we used [Fig.~\ref{fig:Modelling}(b)] and that a different result might be obtained using, for example, a pure sinusoidal waveform as in Ref.~\onlinecite{Fletcher2013}. Using AWG waveforms to drive the pump gives the possibility of manipulating the electron wave packet. In the following sections, we demonstrate such a technique with the pump operated as a two-electron source.

\section{Two-electron pumping}
\begin{figure}
\centering
\includegraphics[width=0.5\textwidth]{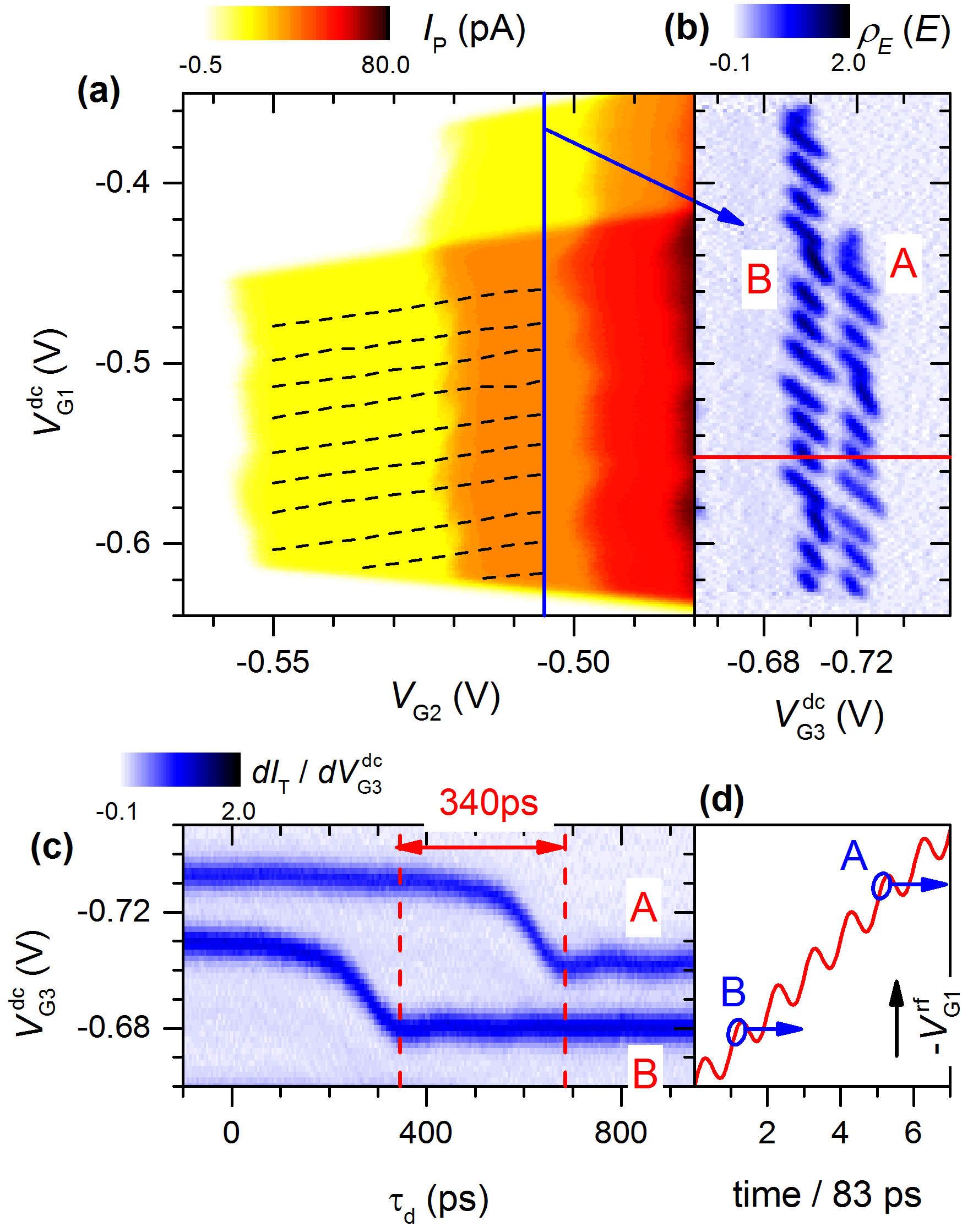}
\caption{(a) \Ip, as a function of \VGoneDC\ and \VGtwo. Dashed lines are contours of constant electron emission time. (b) $\rho_E(E)$ as a function of \VGoneDC\ for two-electron pumping, at \VGtwo\ $=-0.505$~V [solid vertical line in (a)]. (c) \dIbydV\ as a function of \tauD\ for $\text{\VGoneDC}\ = -0.552$~V [solid horizontal line in (b)]. (d) Indicates the relative emission points of electrons A and B in the pumping waveform \VGoneRF. Data taken at $B = 10$~T. (color online)}
\label{fig:TwoElectron}
\end{figure}
The electron pump can be operated as a source of pairs of electrons, by making \VGtwo\ less negative so that two electrons are trapped in the dot and pumped per cycle. Figure \ref{fig:TwoElectron}(a) shows the pumped current as a function of \VGoneDC\ and \VGtwo\ at $B=10$~T, showing clear regions where $\text{\Ip} = nef$ for $n=1,2$ and 3. Before considering the two-electron case, we comment that, since each peak in the electron arrival-time and energy distributions comes from an individual sampling interval of the AWG pumping waveform, we can identify the specific point of emission within the pumping cycle, to within one sampling interval. In Fig.~\ref{fig:EnergyDist}(d), the topmost diagonal line feature is due to electron emission in the sampling interval with the most negative \VGoneRF\ and for each subsequent line the emission moves to earlier time by one sampling interval ($\sim83$~ps). By repeating the map of Fig.~\ref{fig:EnergyDist}(d) at different \VGtwo, we can plot contours of constant emission time, shown as dashed lines in Fig.~\ref{fig:TwoElectron}(a). To our knowledge this is the first measurement of the specific time in the pumping cycle when electron emission occurs.

We note some jitter in the edges of the constant-current regions in the map of \Ip\ in Fig.~\ref{fig:TwoElectron}(a). We believe this jitter is also linked to high-frequency ripples in the AWG pumping waveform. However, the period (in \VGoneDC) of this jitter is different to the period of the dashed lines indicating the emission point. The number of electrons pumped per cycle (and hence \Ip) depends on both the electron capture and electron emission regions of the pumping waveform. We suggest that the jitter in the edges of the constant-\Ip\ regions may be more linked to the high-frequency ripple in the electron capture region, rather than the emission region. This point requires further investigation.

To study two-electron pumping, we set $\text{\VGtwo}=-0.505$~V [vertical line in Fig.~\ref{fig:TwoElectron}(a)]. First we consider the two-electron energy distribution as a function of \VGoneDC, using measurements with a static detector, as shown in Fig.~\ref{fig:TwoElectron}(b). Compared with the single-electron case [Fig.~\ref{fig:EnergyDist}(d)], there are now two sets of diagonal-line features, corresponding to two pumped electrons arriving at the detector barrier with different energies. The higher-energy set of diagonal lines evolves continuously from the single-electron features of Fig.~\ref{fig:EnergyDist}(d) as we make \VGtwo\ less negative. These features are due to the electron that remains in the pump for longest, which we label as electron ``A''. The lower-energy features are due to the electron that leaves the pump first (labelled ``B''). It is noticeable that each diagonal-line feature has a slightly different length and slope, which we link to irregularities in the 12-GHz ripple of the pumping waveform. Looking closely at Fig.~\ref{fig:TwoElectron}(b), we see that the features due to electron B are translated vertically with respect to the features due to electron A by approximately four diagonal-lines, towards less negative \VGoneDC. Recalling that, in the single-electron case, diagonal-lines at more negative \VGoneDC\ are due to electron emission from earlier sampling intervals, this suggests that electron B is emitted four sampling intervals ($\approx330$~ps) earlier than electron A. We attribute this to the increase in electrochemical potential from adding a second electron to the pump, which has a similar effect to making \VGoneDC\ more negative and causes earlier electron emission. We note that Fletcher \textit{et al} observed a similar emission-time gap between two electrons for an electron pump similar to our device. \cite{Fletcher2013} We also find that the emission-time gap is increased to five sampling intervals ($\approx 415$~ps) on increasing the magnetic field from 10 T to 14 T, which may be linked to the effect of the magnetic field on the shape of the bound-state wavefunctions in the pump and hence the tunneling rates.\cite{Fletcher2012}.

For a more accurate measurement of the two-electron time gap, we use the square-wave detector modulation. In Fig.~\ref{fig:TwoElectron}(c) we plot the derivative \dIbydV\ at \VGoneDC\ $=-0.552$~V [horizontal line in Fig.~\ref{fig:TwoElectron}(b)] as a function of the square wave delay \tauD. This data is the two-electron equivalent of Fig.~\ref{fig:TimeDist}(a). We see that electron B arrives at the detector 340~ps before electron A, consistent with our estimate of four sampling intervals, and with energy $\approx12$ meV below the energy of electron A. Fig.~\ref{fig:TwoElectron}(d) shows schematically the relative emission points of the two electrons in the pumping waveform, based on our earlier modeling. The observed energy gap is in contrast to the results of Ubbelohde \textit{et al}, \cite{Ubbelohde2015} who found that for two-electron pumping using a sinusoidal waveform the electrons had equal emission energy, which they attributed to the out-tunneling rate $\Gamma(t)$ depending only on the difference between the energy of the top-most electron and the detector barrier height. However, our result agrees with that of Fletcher \textit{et al}, \cite{Fletcher2013} who found that the first-emitted electron (B) had lower energy and argued that additional factors may enhance the emission rate when two electrons are bound in the pump, so electron B can be emitted with lower energy than in the single-electron case. This may depend on the device geometry, leading to the differing results of Refs. \onlinecite{Fletcher2013} and \onlinecite{Ubbelohde2015}.

\section{Manipulation of two-electron emission}
\begin{figure*}
\centering
\includegraphics[width=0.85\textwidth]{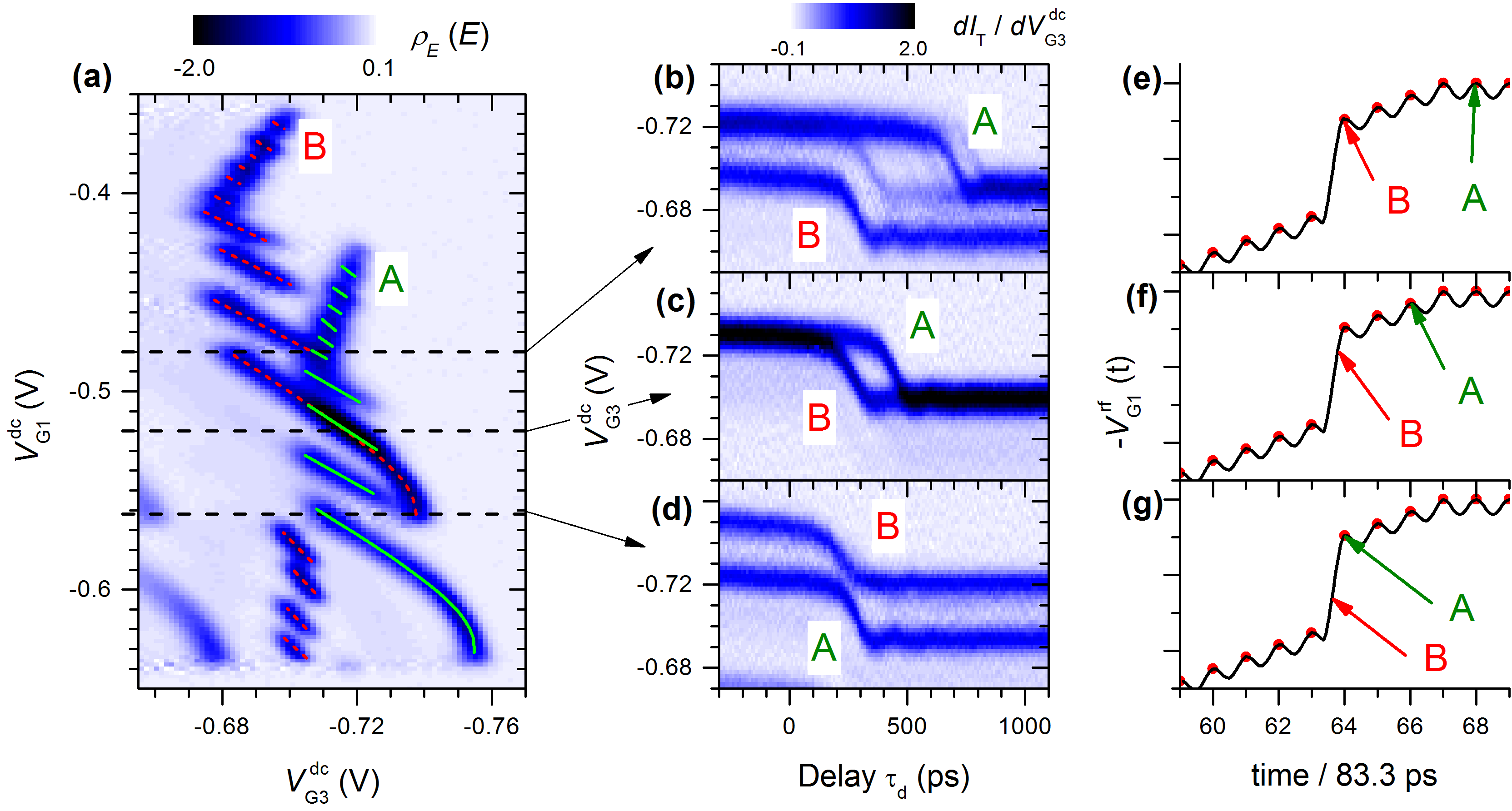}
\caption{(a) $\rho_E(E)$ as a function of \VGoneDC\ using the modified pumping waveform; features due to electron A(B) are coded with green solid (red dashed) lines. (b)-(d) \dIbydV\ as a function of \tauD\ for the three values of \VGoneDC\ shown by dashed lines in (a). (e)-(g) Schematic sketches of the modified pumping waveform, showing the approximate electron emission points for cases (b)-(d). Data taken at $B = 10$~T and $\text{\VGtwo} = -0.505$~V. (color online)}
\label{fig:WFM110}
\end{figure*}
Finally, we show how engineering of the pumping waveform can be used to manipulate the two-electron time and energy gap. We modify the pumping waveform \VGoneRF\ so that the voltage step $\Delta\text{\VGoneRF}$ during one particular sampling interval within the emission region is much larger than for the rest of the steps (large step 128 mV, compared to 16 mV for the other steps). Figure~\ref{fig:WFM110}(e) illustrates this waveform schematically, including the 12-GHz ripple, for the emission part of the pumping cycle. Introducing the large step causes profound changes in the electron emission, as shown in Fig.~\ref{fig:WFM110}(a), where we plot the two-electron energy distribution resulting from the modified pumping waveform as a function of \VGoneDC. Features due to electron A(B) are marked with green solid (red dashed) lines as a guide to the eyes. Compared to pumping with the digital sine wave [Fig.~\ref{fig:TwoElectron}(b)], the main change is that emission during the large-step sampling interval occurs over a much wider interval of \VGoneDC, with a very large increase in energy for the most negative \VGoneDC\ of this interval. We do not attempt to understand this situation quantitatively, because it is not known how the finite-bandwidth signal line will transmit the modified waveform to the sample and because the large step may cause excitation of electrons to higher orbital states of the dot.\cite{Kataoka2011} However, we believe electron emission still occurs by sequential tunneling, rather than ballistically, so we can gain some insight using the tunneling model presented earlier.

The tunneling model predicts that electron emission is generally pinned to the local maximum of $- \text{\VGoneRF}$ in one particular AWG sampling interval, so that the emission energy rises linearly as \VGoneDC\ is made more negative, until emission from the local maximum in the preceding sampling interval becomes possible. For the large-step waveform, a very large negative shift in \VGoneDC\ is required to shift emission from the local maximum at the top of the large step to the local maximum of the previous sampling interval. Therefore emission is pinned to the top of the large step for a wide range of \VGoneDC. The total voltage $\text{\VGoneDC}+\text{\VGoneRF}$ at the top of the large step increases linearly with increasingly negative \VGoneDC, pushing up the emission energy. However, this effect seems to saturate, shown by the longest diagonal line features in Fig.~\ref{fig:WFM110}(a) becoming curved for the most negative \VGoneDC, perhaps because the emission rate becomes fast enough for emission on the riser of the large step, before the local maximum.

Figure~\ref{fig:WFM110}(a) shows that with the modified pumping waveform it is possible for the energy gap $E_\text{A}-E_\text{B}$ to be reduced and even reversed. Once again, we use the square-wave detector modulation to reveal the details of the two-electron arrival-time difference and energy gap, with the results shown in Fig.~\ref{fig:WFM110}(b)-(d) for three values of \VGoneDC. In Fig.~\ref{fig:WFM110}(b), both electrons are emitted after the large step in the pumping waveform so the situation is the same as in Fig.~\ref{fig:TwoElectron}(c), with the electrons emitted four sampling intervals apart, and $E_\text{A}-E_\text{B}\approx12$ meV. As we make \VGoneDC\ more negative, emission of electron A is pushed to earlier sampling intervals but emission of electron B stays fixed on the large step, with a consequent increase in $E_\text{B}$. This makes it possible for the two electrons to be emitted with equal energies only $\approx200$~ps apart [Fig.~\ref{fig:WFM110}(c)], or for the two electrons to be emitted in the same sampling interval (time gap $\approx60$~ps) with reversed energy gap $E_\text{B}-E_\text{A}\approx13$ meV [Fig.~\ref{fig:WFM110}(d)]. Figures \ref{fig:WFM110}(e)-(g) show the approximate emission points for the two electrons, based on the emission times from Fig.~\ref{fig:WFM110}(b)-(d). These results demonstrate the potential of this technique to control the two-electron wave packet.

\section{Conclusion}
In conclusion, we have measured the arrival-time and energy distributions of single electrons and pairs of electrons emitted from a semiconductor electron pump, with sufficient time-resolution to determine an upper bound of 30~ps FWHM for the width of the single-electron arrival-time distribution. Our measurement technique has the potential for even further improvement in time-resolution, by increasing the rate of the detector modulation and by reducing cross-talk between the pump and the detector. We have shown how the details of the waveform used to drive the electron pump affect the electron emission, in agreement with a tunneling model of the emission process. This enables manipulation of the electron time and energy distributions, which was demonstrated using the example of controlling the time difference and energy gap between a pair of electrons. Measurement and control of the two-electron time and energy gap could be particularly useful when combined with measurements of the electron partitioning noise,\cite{Ubbelohde2015} for studying the factors that determine the degree of correlation within the emitted electron pairs. Our results highlight the potential of the semiconductor electron pump as an on-demand emitter of single electrons and pairs of electrons, with fine control of the emission time, energy and wave packet shape. We expect this to find application in studies of fermionic quantum behavior and preparation of electron states for quantum information processing.

\begin{acknowledgments}
We thank J.D.~Fletcher, S.P.~Giblin and D.A.~Humphreys for useful discussions and C.A.~Nicoll and R.D.~Hall for technical assistance. This research was supported by the UK Department for Business, Innovation and Skills, NPL's Strategic Research Programme and the UK EPSRC. V.K.~has been supported by the Latvian Council of Science within research project no.~146/2012.
\end{acknowledgments}

% If in two-column mode, this environment will change to single-column
% format so that long equations can be displayed. Use
%\begin{widetext}
%\end{widetext}

% figures should be put into the text as floats.
% Use the graphics or graphicx packages (distributed with LaTeX2e)
% and the \includegraphics macro defined in those packages.
% See the LaTeX Graphics Companion by Michel Goosens, Sebastian Rahtz,
% entire page. There is no need to do explicit centering.

% \begin{figure}
% \end{figure}

% Surround figure environment with turnpage environment for landscape
% figure
% \begin{figure}
% \end{figure}

%\begin{acknowledgments}
%\end{acknowledgments}

%\bibliography{QDEBreferences}

\end{document}